\documentclass[aps,prl,showpacs,nofootinbib,floatfix,twocolumn]{revtex4}

\usepackage{bm}
\usepackage{latexsym}
\usepackage{dcolumn}
\usepackage{amsfonts,amssymb,amsmath}
\usepackage{graphicx,epsfig}
\usepackage{psfrag}
\usepackage{subfigure}
\usepackage{feynmf}
\usepackage{rotating}
\usepackage{hyperref}
\usepackage{color}

\hypersetup{
    bookmarks=true,         
    unicode=false,          
    pdftoolbar=true,        
    pdfmenubar=true,        
    pdffitwindow=false,     
    pdfstartview={FitH},    
    pdftitle={My title},    
    pdfauthor={Author},     
    pdfsubject={Subject},   
    pdfcreator={Creator},   
    pdfproducer={Producer}, 
    pdfkeywords={keyword1} {key2} {key3}, 
    pdfnewwindow=true,      
    colorlinks=true,       
    linkcolor=red,          
    citecolor=cyan,        
    filecolor=magenta,      
    urlcolor=green,           
    linktocpage=true
}

\def\beq{\begin{equation}}
\def\eeq{\end{equation}}
\def\br{\begin{eqnarray}}
\def\er{\end{eqnarray}}
\def\benu{\begin{enumerate}}
\def\efnu{\end{enumerate}}
\def\nn{\nonumber}

\def\l{\left}
\def\r{\right}

\def\cl{{C}_{\ell}}

\begin{document}
\title{Cosmological parameter estimation with free-form primordial power spectrum}
\author{Dhiraj Kumar Hazra}\email{dhiraj@apctp.org} 
\affiliation{Asia Pacific Center for Theoretical Physics, Pohang, Gyeongbuk 790-784, Korea}
\author{Arman Shafieloo}\email{arman@apctp.org}
\affiliation{Asia Pacific Center for Theoretical Physics, Pohang, Gyeongbuk 790-784, Korea\\ Department of Physics, POSTECH, Pohang, Gyeongbuk 790-784, Korea}
\author{Tarun Souradeep}\email{tarun@iucaa.ernet.in}
\affiliation{Inter-University Centre for Astronomy and Astrophysics, Post Bag 4,
Ganeshkhind, Pune 411~007, India}
\begin{abstract}
Constraints on the main cosmological parameters 
using CMB or large scale structure data are usually based on power-law assumption of the primordial power spectrum (PPS). 
However, in the absence of a preferred model for the early universe, this raises a concern that current cosmological 
parameter estimates are strongly prejudiced by the assumed power-law form of PPS. In this paper, for the first time, we 
perform cosmological parameter estimation allowing the free form of the primordial spectrum. This is in fact the most general
approach to estimate cosmological parameters without assuming any particular form for the primordial spectrum. We use 
direct reconstruction of the PPS for any point in the cosmological parameter space using recently modified Richardson-Lucy
algorithm however other alternative reconstruction methods could be used for this purpose as well. We use WMAP 9 year data in our analysis considering CMB lensing effect and we report, for the first time, that the 
flat spatial universe with no cosmological constant is ruled out by more than $4\sigma$ confidence limit without assuming any 
particular form of the primordial spectrum. This would be probably the most robust indication for dark energy using CMB data 
alone. Our results on the estimated cosmological parameters show that higher values of baryonic and matter density and lower value of Hubble parameter (in comparison 
to the estimated values by assuming power-law PPS) is preferred by the data. However, the estimated cosmological parameters by 
assuming free form of the PPS  have overlap at $1\sigma$ confidence level with the estimated values assuming power-law form of PPS.      

\end{abstract}
\maketitle

\section{Introduction}~The observables of the perturbed universe, such as CMB anisotropy, galaxy surveys and weak lensing, all depend on a set of
cosmological parameters assuming a background model describing the current universe, as well as the parameters characterizing the
presumed nature of the initial perturbations. While certain characteristics of the initial perturbations, such as the adiabatic 
nature and tensor contribution, can and are being tested independently, the shape of the PPS remains, at best, a well-motivated 
assumption. It is important to distinguish between the cosmological parameters within a model that describe the present universe
from that characterizing the initial conditions, specifically the PPS, $P(k)$. The standard model of cosmology which is the most 
popular and widely used cosmological model is the spatially flat $\Lambda$CDM model which incorporates a power law form of the 
primordial power spectrum. The model is described by 6 parameters. Four of them describe the background $\Lambda$CDM, represented 
by 
$\Omega_b$ (\rm baryon density), $\Omega_{\rm CDM}$ (cold dark matter density), $H_0$ (present rate of expansion of 
the universe)\footnote{Sometimes ratio of the sound horizon to the angular diameter distance at decoupling, 
$\theta$ is considered to be a parameter instead of $H_0$} and the reionization optical depth $\tau$. 
We should mention, dark energy density $\Omega_{\rm \Lambda}$ is directly obtained as a remainder of baryon and 
cold dark matter density from the total density as we have assumed spatially flat model of the universe. The other two parameters
in the model describe the form of the primordial power spectrum which is assumed to be the power law defined
by $P(k) = A_s[\frac{k}{k_{\ast}}]^{ns-1}$ where ${A}_{s}$ is the amplitude\footnote{Note that this amplitude is 
defined at some pivot scale $k_{\ast}$} and the tilt is given by the spectral index $n_{\rm S}$. The imposed form 
of the primordial spectrum allows us to provide tight constraints on the four background parameters, however, these
tight constraints are basically the result of the rigidness of the model and certain assumption of the primordial spectrum.
In other words choosing different assumptions for the form of the primordial spectrum result in different constraints on 
the background cosmological parameters~\cite{Shafieloo:2009cq}. In this paper, for the first time we study the complete Markov 
Chain Monte Carlo parameter estimation assuming a free form of the primordial spectrum through direct reconstruction of the PPS for 
each point in the background cosmological parameter space using WMAP 9 year data~\cite{wmap-lambda,Hinshaw:2012fq}. There have been 
different interesting attempts to directly reconstruct the form of the primordial 
spectrum~\cite{Shafieloo:2009cq,Hannestad:2000pm,Tegmark:2002cy,Shafieloo:2003gf,Bridle:2003sa,Mukherjee:2003ag,Hannestad:2003zs,
TocchiniValentini:2005ja,Kogo:2005qi,Leach:2005av,Shafieloo:2006hs,Shafieloo:2007tk,Nagata:2008tk,Nagata:2008zj,Souradeep:2008zz,
Ichiki:2009zz,Paykari:2009ac,Nicholson:2009pi,Nicholson:2009zj,bridges,Gauthier:2012aq,Hlozek:2011pc,ppsrecon} and 
in our analysis we use the recently modified Richardson-Lucy algorithm~\cite{ppsrecon}. We show that it is indeed possible to do the
cosmological parameter estimation allowing free form of the primordial spectrum and for the first time we report that a spatially flat 
universe without cosmological constant is ruled out by more than $4\sigma$ confidence limit using CMB data alone. This is without putting any prior 
constraints on the Hubble parameter or using any other cosmological observation. We show that assuming a free form primordial 
spectrum, the confidence limits of the background cosmological parameters are larger, as expected than those we get by assuming the
power-law form of the PPS and we present that the data prefers larger values of baryonic and matter densities for the free form of the 
primordial spectrum in comparison with power-law assumption. In the next section we discuss the methodology of reconstruction and the 
parameter estimation followed by which we shall demonstrate our results in the results section. We close with a brief discussion at the end.
It should be noted that the aim of this paper is not to relax a parameter of an underlying cosmological model and 
investigate the effects on other cosmological parameters as usually done to study the cosmographical degeneracies. In this paper we 
perform cosmological parameter estimation analysis with free form primordial power spectrum which directly 
indicates that we do not consider any assumptions on underlying inflationary models(or any theoretical model of the early universe).
We have been able to compare the free form spectra because of our method~\cite{ppsrecon} is able to identify 
PPS functions with very large improvement to the WMAP likelihood at any point in cosmological parameter space.
This allows estimation of cosmological parameters optimized over the PPS functional degree of freedom.

\section{Formalism}
In this work, we have used the recently modified version of Richardson-Lucy algorithm ~\cite{ppsrecon} (we 
call it here after MRL) 
to reconstruct the optimal form of the primordial spectrum for each point in the background cosmological parameter space. The 
Richardson-Lucy algorithm~\cite{richardson,lucy,baugh1,baugh2} has been used previously in this context to reconstruct the 
primordial spectrum ~\cite{Shafieloo:2003gf,Shafieloo:2006hs,Shafieloo:2007tk} and in the recently modified version one can 
use the combination of un-binned and binned data in the analysis~\cite{ppsrecon}. The modified algorithm can be formulated as,

\begin{widetext}
\begin{eqnarray}
{{P}_{k}^{(i+1)}}-{{P}_{k}^{(i)}}&=&{{P}_{k}^{(i)}}\times\Biggl[\sum_{\ell=2}^{\ell=900}
{\widetilde{G}}_{\ell k}^{\rm un-binned}\Biggl\{\l(\frac{\cl^{\rm D}-\cl^{{\rm T}(i)}}{\cl^{{\rm T}(i)}}\r)~\tanh^{2}
\l[Q_{\ell} (\cl^{\rm D}-\cl^{{\rm T}(i)})\r]\Biggr\}_{\rm un-binned}\,\nn\\
&+& \sum_{\ell_{\rm binned}>900}
{\widetilde{G}}_{\ell k}^{\rm binned}\Biggl\{\l(\frac{\cl^{\rm D}-\cl^{{\rm T}(i)}}{\cl^{{\rm T}(i)}}\r)~\tanh^{2}
\l[\frac{\cl^{\rm D}-\cl^{{\rm T}(i)}}{\sigma_{\ell}^{\rm D}}\r]^{2}\Biggr\}_{\rm binned}\Biggr],~\label{eq:combined}
\end{eqnarray}
\end{widetext}

Where, ${{P}_{k}^{(i+1)}}$ and ${{P}_{k}^{(i)}}$ are the power spectrum evaluated in iterations $i+1$ and $i$
respectively and the quantity ${\widetilde{G}}_{\ell k}$ is normalized radiative transport kernel ${G}_{\ell k}$. $\cl^{\rm D}$
and $\sigma_{\ell}^{\rm D}$ are the observed $\cl^{\rm TT}$ data points and the corresponding diagonal terms of the inverse 
covariance matrix. $\cl^{{\rm T}(i)}$ is theoretically calculated angular power spectrum in $i$'th iteration. $Q_{\ell}$ is given by the following expression,
\beq
Q_{\ell}=\sum_{\ell'}(C_{\ell'}^{\rm D}-C_{\ell'}^{{\rm T}(i)}) COV^{-1}(\ell,\ell'),
\eeq

where $COV^{-1}(\ell,\ell')$ is the inverse of the error covariance matrix. 

The radiative transport kernel, ${G}_{\ell k}$ which depends on the background cosmological parameters, satisfies the following equation. 
\beq
\cl=\sum_{i}{G}_{\ell k_{i}}{{P}_{k_{i}}}~\label{eq:clequation}
\eeq

We should mention that as has been indicated in eq.~\ref{eq:combined} we have used the un-binned data till $\ell=900$ and used the 
binned data thereafter because of the increasing noise in higher multipoles. For more discussion on this procedure see, 
ref.~\cite{ppsrecon}.

We use publicly available CAMB~\cite{cambsite,Lewis:1999bs} to calculate the kernel and the $\cl$'s and CosmoMC~\cite{cosmomcsite,Lewis:2002ah} 
to perform the Markov Chain Monte Carlo (MCMC) analysis on the cosmological parameters. 
 Most recent WMAP nine-year observational data~\cite{Hinshaw:2012fq} has
been used in this analysis. We have taken into account the effect of gravitational lensing through approximating the lensing effect on different background models.
Basically we assume that for each point in the background cosmological parameter space the lensing effect on the observed angular power 
spectrum would be the same if we assume the power-law PPS or we use the reconstructed PPS. 
To do so, for each point in the background parameter space one can identify the contribution of gravitational lensing 
to $\cl^{\rm TT}$'s for the best fit power law primordial spectrum from computed lensing potential 
power spectrum and using curved-sky correlation function method. We subtract the contribution from the 
WMAP-unbinned temperature anisotropy data and follow the reconstruction from the modified data. Finally from the 
reconstructed primordial spectrum we calculate the CMB temperature and polarization $\cl$'s and we lens all the $\cl$'s again 
using curved-sky correlation through CAMB~\cite{cambsite,Lewis:1999bs} and compare them with WMAP data.  
We should mention that for the multipole range covered by WMAP, the effect of lensing on the temperature and polarization 
spectrum is not substantial and we find that without lensing too we recover our conclusions of this paper. In order to
have a complete analysis we have included the effect of lensing.


We have performed the MRL up to 40 iterations. However we have checked the consistency of our results allowing MRL to work up to different 
iteration numbers. 


\begin{figure*}[!htb]
\begin{center}

\includegraphics[width=150mm]{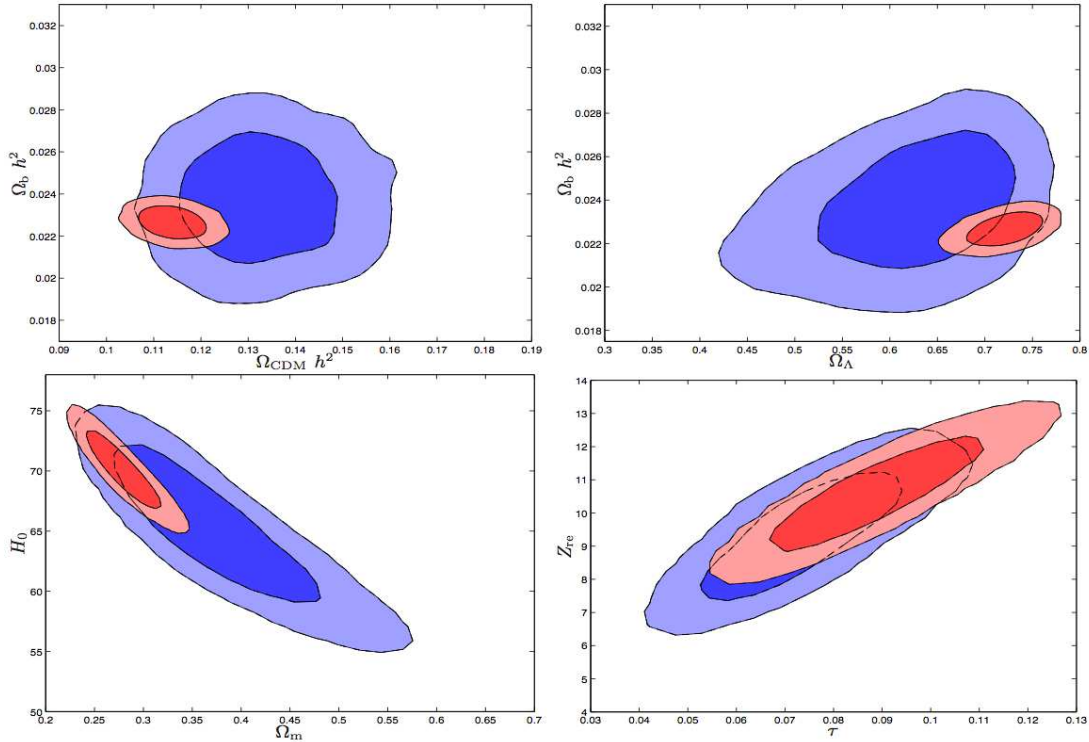}

\end{center}
\caption{\footnotesize\label{fig:contours} 2-dimensional confidence contours of the background cosmological parameters. The blue contours represents the results of the 
analysis allowing free form of the primordial spectrum while the red contours are by assuming the power law form of the primordial spectrum. As expected the free form of the 
primordial spectrum relaxes the bounds on the parameters. Throughout the analysis we have assumed a spatially flat universe and one can see that a universe with zero density 
for the cosmological constant is ruled out with high confidence even with no assumption for the primordial spectrum. In comparison with the results from power law assumption 
of the primordial spectrum, the data prefers higher values of baryonic and matter densities when we allow free form of the primordial spectrum.}
\end{figure*}
\section{Results}~\label{sec:results}
Following the methodology described in the previous section, here we shall discuss the result of our MCMC analysis with the nine year data from WMAP.
We find the best fit model provides a $\chi^2$ ($-2~\ln {\cal L}$) of 7441.4 which is about 115 better than what we get from the best fit power law spectrum. 
We compare the bounds on the background cosmological parameters with the power law results in figure~\ref{fig:contours}. 

\begin{figure}[!htb]
\begin{center} 
\includegraphics[angle=270,width=90mm]{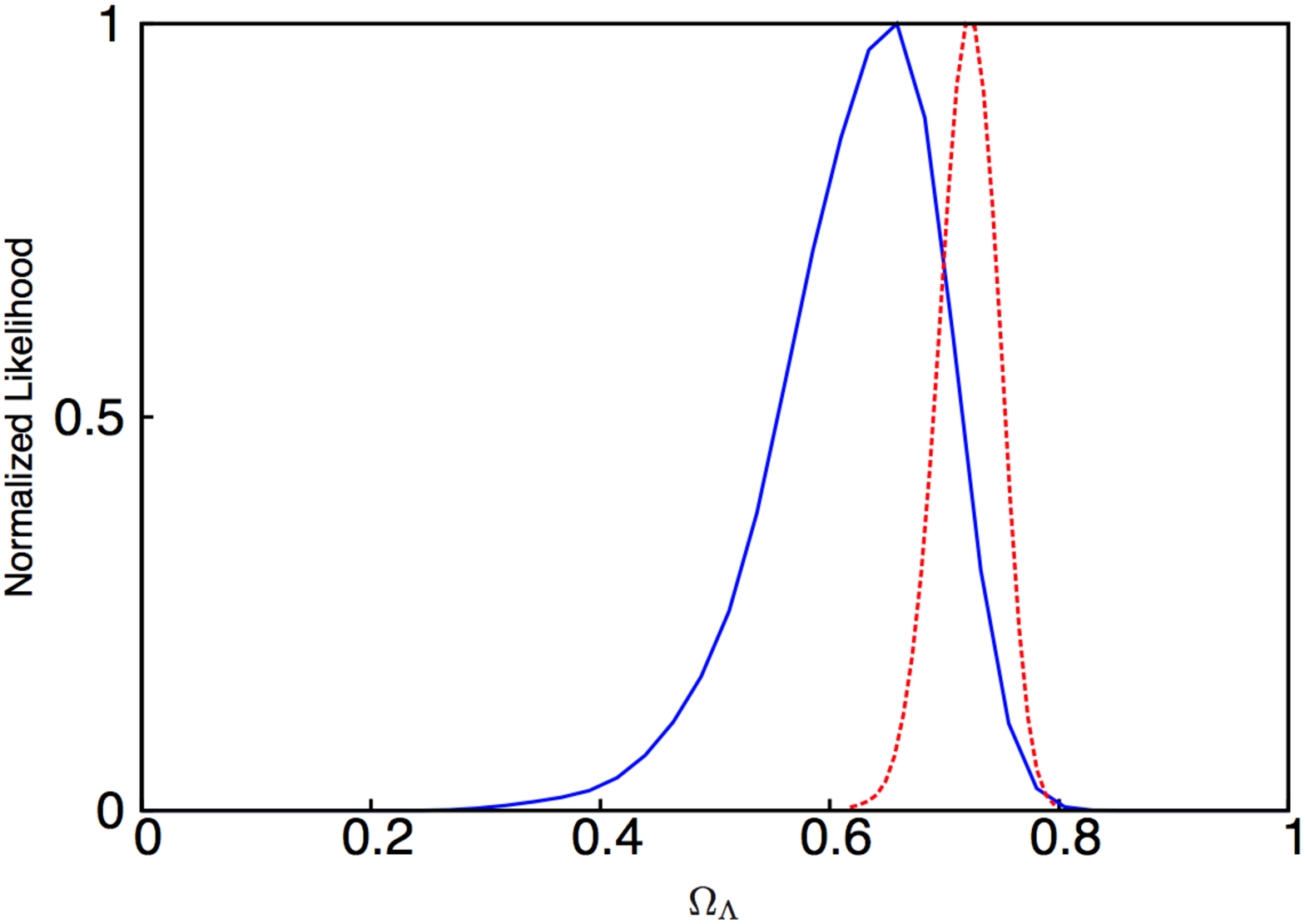}
\end{center}
\caption{\footnotesize\label{fig:de1d} The one dimensional marginalized likelihood of dark energy density $\Omega_\Lambda$ obtained using free form of primordial
spectrum (in solid blue line) and using power law (in dashed red line). $\Omega_\Lambda=0$ is clearly not favored by the data even if we allow a power spectra free of forms. Quantitatively, in 4$\sigma$ the data rules out $\Omega_\Lambda<0.25$. This is probably the first indication towards presence of dark energy with a very high confidence using CMB data alone.}
\end{figure}

As we are allowing a free form primordial power spectrum, hence having larger effective degrees of freedom, it is expected to have bigger 
confidence contours compared to the confidence limits derived by assuming power law PPS and fig.~\ref{fig:contours} clearly illustrates this fact. 
The results from Fig.~\ref{fig:contours} suggest that allowing free form PPS, data prefers higher baryon and dark matter densities compared to the 
canonical results. Best fit values of $\Omega_{\rm b}h^2,~\Omega_{\rm CDM}h^2,~{\rm H}_0$ and $\tau$ are $0.0232$, $0.132$, $64$ and $0.077$ respectively. 
It should be pointed out that our analysis allows a lower value of optical depth ($\tau$) 
which, in turn allows low redshift of reionization ($Z_{\rm re}\sim7$) compared 
to the power law results (see last plot of Fig.~\ref{fig:contours}).
Our results also indicates that independent of the form of the PPS, 
a flat model of universe with no cosmological constant is ruled out with a very high confidence. To our knowledge this is the first direct indication 
towards dark energy with high certainty from CMB temperature and polarization data analysis alone assuming spatial flatness. In figure~\ref{fig:de1d} we plot the one 
dimensional marginalized likelihoods of the parameter $\Omega_\Lambda$ obtained using power law (in dashed red line) and allowing free form of the 
primordial spectrum. The plot clearly demonstrates that a low value of dark energy density is ruled out. Obtained result suggests that 
values of $\Omega_\Lambda < 0.25$ is ruled out at 4$\sigma$ which implies a strong exclusion of $\Omega_\Lambda =0$ with a very high confidence. In this
context it should be noted that no-dark-energy model including curvature was previously ruled out by 3.2-$\sigma$ using Atacama Cosmology Telescope lensing 
measurements~\cite{Sherwin:2011gv} but within the assumption of a power law form of PPS.

\begin{figure}[!htb]
\begin{center} 
\includegraphics[angle=270,width=90mm]{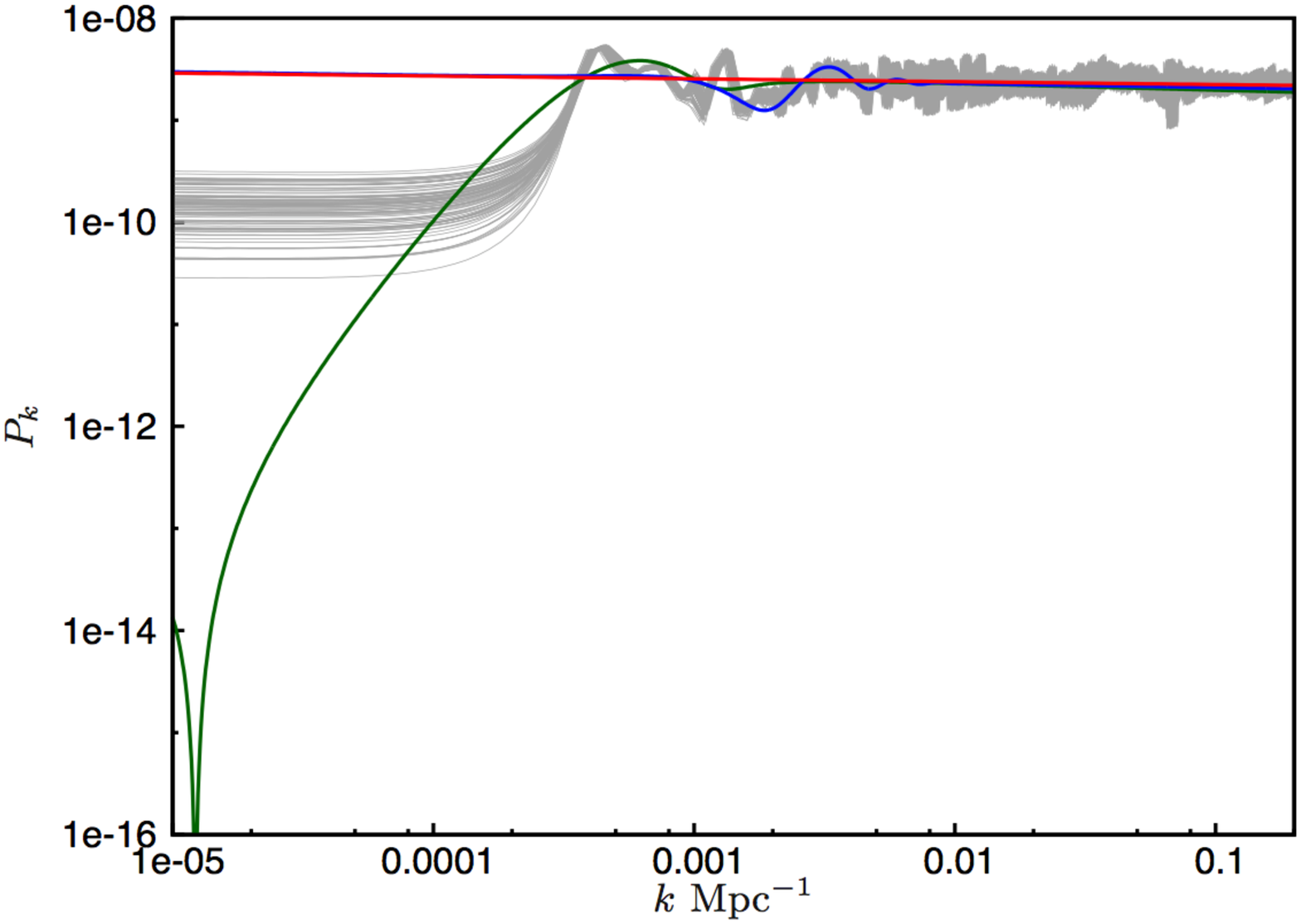}
\end{center}
\caption{\footnotesize\label{fig:ps-error} Reconstructed power spectrum (in grey) obtained from parameters lying within the 2$\sigma$ range 
of the best-fit likelihood. Over the sample of the reconstructed
spectra we have plotted the best fit spectra from the step model~\cite{step} (in blue) and the punctuated model~\cite{puncinf} (in green) of inflation. The best fit power law power spectrum (in red) is 
plotted as well for comparison. Note that barring the low-$k$ region (which data has low sensitivity) the sample of reconstructed spectra 
incorporates nearly all the models within its 2$\sigma$ variation.}
\end{figure}

In Fig.~\ref{fig:ps-error} we plot a few (nearly 100) power spectra (in grey) reconstructed from the WMAP nine-year data with the kernels
corresponding to the cosmological parameters within
$95\%$ (2$\sigma$) limits of the best fit. We also show the best fit power spectrum from the punctuated inflation model~\cite{puncinf} in green 
and the step models of 
inflation in blue~\cite{step} for comparison. The best fit power law power spectrum is also plotted (in red). We should mention 
that while in the reconstruction process we have used only temperature data, in the likelihood analysis polarization data is included. 
As discussed in~\cite{ppsrecon} considering the WMAP polarization data does not significantly improve the reconstruction procedure due to 
low quality of the polarization data.   

We should mention that to check the validity of our approximation regarding lensing contribution, we have repeated our analysis without subtracting 
the power law lensing effects from the data and we find that the later comparison provides a $\chi^2$ worse by 6 than the actual analysis which in turn 
indicates that our approximation on lensing contribution works well.  

 To check the robustness of our analysis and the validity of the obtained results, we performed some tests with simulated data. We have synthesized number of  $\cl^{\rm TT}$ data from angular power spectrum obtained using power law and $\Lambda$CDM model with some fixed parameters. 
 With the reconstructed free form spectrum we perform MCMC on the following datasets and calculated the likelihood assuming $\chi^2$ distribution~\footnote{We have shown recently~\cite{ppsrecon} 
that this likelihood estimator is robust and can be used as an approximation to the complete WMAP likelihood}. 
We find that in most of the cases (more than 90\%) the obtained confidence 
contours of the cosmological parameter contains the fiducial parameter values within the 2$\sigma$ region. This indicates, with the reconstruction we get our 
fiducial model back in almost all the cases, which, in turn proves the robustness of our analysis.

\section{Discussion}~\label{sec:discussion}
In this paper we have estimated the cosmological parameters assuming free form of
the primordial spectrum. The primordial spectrum for each point 
in the background cosmological parameter space is obtained using the MRL reconstruction procedure using the WMAP nine-year 
combined data of un-binned
and binned angular power spectrum. We should mention that for this analysis and instead of MRL one can use trivially other 
alternative methods of non-parametric reconstruction 
of the primordial spectrum. In fact the MRL method serves just as a possible method of reconstruction to get a PPS that improves the fit of a 
cosmological model at different points in the cosmological parameter space to the CMB data. One is free to choose any other method to do this task.
The background model is assumed to be a spatially flat $\Lambda$CDM model. Performing the MCMC analysis using  
CosmoMC we obtained the bounds on the background parameters and we find out that the data prefers higher baryonic and matter densities 
(hence lower $\Omega_{\Lambda}$) and lower Hubble parameter when we assume 
the free form of the PPS in comparison with the case of power-law assumption. We should mention here earlier 
efforts~\cite{Hlozek:2011pc,Blanchard:2003du,Hunt:2007dn} have indicated that allowing deviation from simple power law PPS prefers a lower
Hubble constant and our result too agrees with that. However, with the ever increasing quality of CMB data from WMAP, we find, our result does not
agree with~\cite{Blanchard:2003du} anymore, where it has been shown that allowing deviations from the power law PPS, zero dark-energy model fits the data 
as well as the $\Lambda$CDM model. 
Our results indicates that independent of the form of the primordial spectrum 
and without any prior on the value of the Hubble parameter, the spatially flat universe with no cosmological constant is ruled out with a
very high confidence using WMAP 9 year data alone. This is the first direct evidence of dark energy with a very high certainty from 
CMB data alone and with no prior on the Hubble parameter or assuming the form of the primordial spectrum. We expect to get tighter 
constraints on the background parameters assuming the free form of the primordial spectrum using upcoming Planck data~\cite{planck}.



{\it \bf Acknowledgments:~}We would like to thank Teppei Okumura for useful discussions. D.K.H and A.S wish to acknowledge support from the Korea Ministry of Education, Science
and Technology, Gyeongsangbuk-Do and Pohang City for Independent Junior Research Groups at 
the Asia Pacific Center for Theoretical Physics. We also acknowledge the use of publicly 
available CAMB and CosmoMC to calculate the radiative transport kernel and the angular power spectra.
D.K.H would like to acknowledge the use of the high performance computing 
facilities at the Harish-Chandra Research Institute, Allahabad, India
({\tt http://cluster.hri.res.in}). TS acknowledges support from Swarnajayanti 
fellowship grant of DST, India. 

\end{document}